Andrzej Jarynowski

*Moldova State University in Kishinev/ Smoluchowski Institute, Jagiellonian University in Cracow*

# HPV and cervical cancer in Moldova – epidemiological model with intervention's cost – benefit/effectiveness analysis

The Republic of Moldova has recently started battle with sexual health problems, but due to financial crisis, lack of program coordination, partly because of a lack of experience, sexually transmitted infections have still high morbidity and mortality. From latest ECDC reports Moldova is an European country with highest cervical cancer risk s classified by WHO in the field of sexually transmitted diseases at the level of countries of the third world. Additionally political transformation and beginning of social norm change in Moldova increase sexuality patterns – number of sexual partners (second demographic transition or post-modernist revolution). Computer simulations would answer some questions: when and with what probability should we expect a wave of such epidemic in Moldova. We would like to model multi-faceted transmission of diseases in the context of determining the best epidemiological control. There is also significant difference between urban and rural patterns (in terms of every important aspects of this study: demographic structure, health service access and sexuality). Key questions are to systematically collect and summarize relevant and current data on:

- HPV infections patterns, with type-specific HPV incidence, prevalence and clearance rates

- The optimal preventive guidelines: cervical screening practice, targeted vaccination and sexual education

- Costs of HPV vaccination, screening, treatment and preventive program.

Science (as physics, mathematics and computer science) should take the responsibility of enhancing preventive health prevention strategies. We want decision makers to be informed about proposition of changes to be made in the allocation of the health resources required to implement optimal (cost/effective) prevention program.

*Acknowlagments:* Analysis was financed by EU project EM "IANUS". Author thanks F. Paladi, G Gubceac and E. Cernov for discussions

**Table of context**

# Introduction of HPV/cervical cancer analysis

Human papillomavirus, or HPV, is a sexually transmittable virus infection, which is not only the main, but also necessary risk factor for developing cervical cancer - first most common type of cancer in working age women in Moldova. The time between getting infected by HPV and developing a cancer can be twenty years or more, therefore a dynamic model of human behavior would be very useful, so that simulations can be made and different scenarios compared. We observe both behavioral change (sexuality increase) and demographical change (population ageing). Among the oncogenic HPVs, the most severe one is type 16, present in about half of all cervical cancer cases-we model one strain (16) and imitate multi-strains environment. A lot of medical data are used as model parameters. Recent studies have shown that the main safety precaution with respect to cervical cancer is going to be a combination of vaccination and screening - since only type specific vaccines (for type 16 and 18) are available and there are as many as 15 high risk HPVs.

The main goal is the authoritative analysis of the costs and losses of potential epidemiological control strategies and identifies potential problems that health care will have to face in the future,

● Very little data about sexual contacts (sensitive data),

● There was no verified so far, working model of STI sexually transmitted infections dedicated to the Moldavian community.

## Cervical cancer in Moldova – some fact sheet assumption to model

The Republic of Moldova is one of Europe's poorest nations. Total expenditure on health amounts to just 150EUR per capita (10 times less than in Poland for example). The economic situation over the past 2 decades has not allowed for health systems development and reliable data on cervical cancer were missing. Till now some statistics differ significantly dependently of data source. Demographically, Republic of Moldova had (in years 1998-2014) a population of 1.40-1.60 millions women ages 15 years and older who are at risk of developing cervical cancer. Sexual active woman cohort (age 15-65) was in range 1.25-1.35 millions. Last 10 years estimates indicate that every year 400-550 (approx data) women are diagnosed with cervical cancer and 145-220 (register data) die from the disease. Estimated society cost is around 6000 disability-adjusted life years (DALYs) every year (or 4000 QALY). Cervical cancer ranks as the 3rd most frequent cancer among women in Republic of Moldova and the 1st most frequent cancer among women between 15 and 44 years of age. The incidence of cervical cancer, had increased from 2005 to 2009. It is 39.3% of all kind of woman cancer cases. Data on the HPV burden is not available, but generalizing old studies with data from other south-eastern Europe countries 15-20 % of women in the general population have HPV and 80-90% had it in their live. Prevalence of oncogenic HPV-16/18 is estimated on 2-5%.

### Fatal cases registry

The only reliable data on cervical cancer in Moldova are death registry. However, provided model can predicted new cancer cases, so incidence is needed. 5 years Survival rates decease in last 10 years from 70.4% in 2000 to 61.5% in 2011. Recent numbers in 2013: 2394 persons (58.7%) and in 2012: 2487 (58.8%).

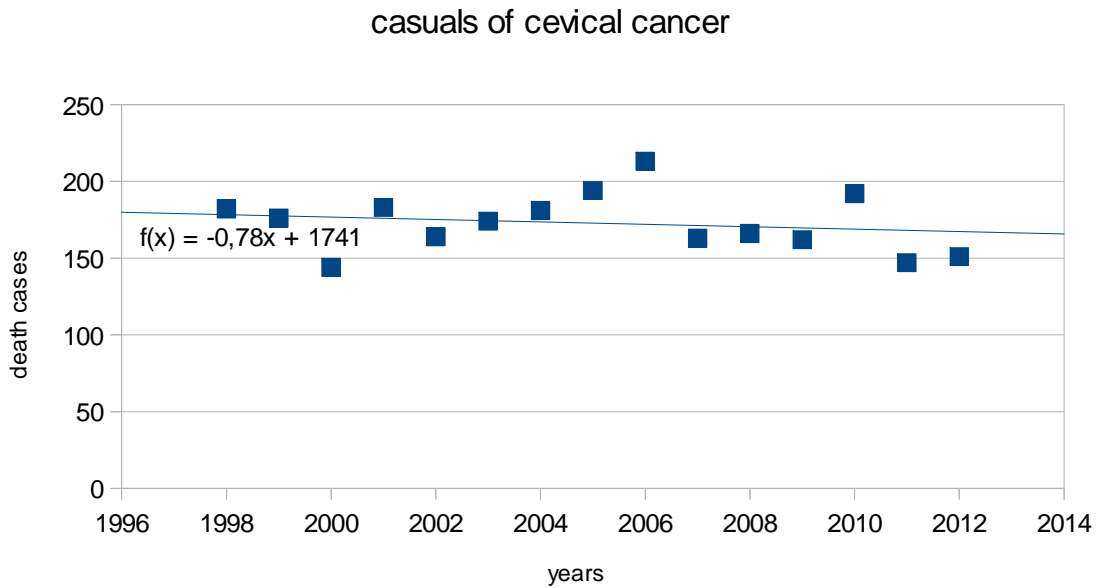

**Figure 1 Death cases registry**

Fatal cases are not good estimators of incidence, because death is shifted in time from caner discovery, and mortality rates increase in time (over 10% in investigated period). Recently, in 2013 there were 198 cases, but in 2012 only 165. Mortality also depend on age (survival chance is decreasing with age).

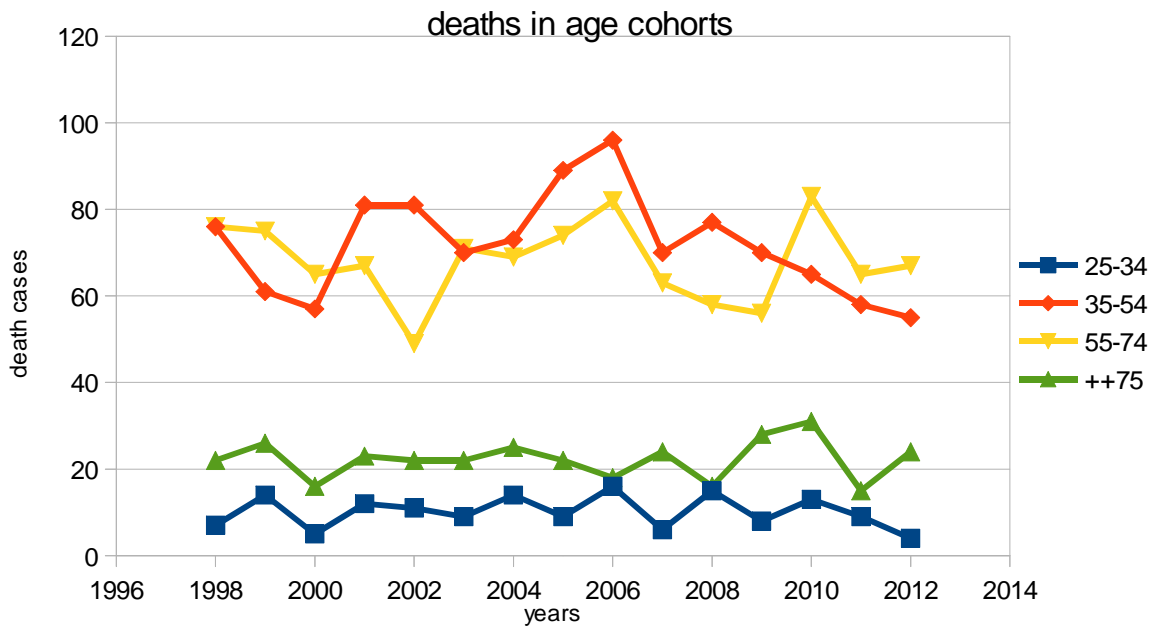

**Figure 2 Death cases in age cohorts**

## Incidence

According do incomplete incidence estimation, we have reliable statistics since 2012, where 475 cases where register, but only 319 was categorized by age:

**Table 1 Confirmed case registry**

| Age | Cases |
|---|---|
| 20-24 | 3 |
| 25-34 | 21 |
| 35-64 | 248 |
| 65++ | 47 |

Base on such a incomplete information (we have detailed registered data only since 2012 and 30% of them were not categorized properly), we could only estimate magnitude of incidence, but it already shows some difference with western countries. As it was indicated, incidence rate was growing last decade (probably partly caused by huge development in surveillance and diagnosing system), while death cases were decreasing the same time (mostly due to better treatment).

## Demographics

In our model we are interesting in sexual active (15-65 years old) part of population (with tracing older woman with HPV, who could develop cancer later on). Mentioned population was growing by last 15 years till now, but because of negative natural growth rates it will suppose to decay in next 15 years. Society is ageing since around 2004 (indicate cancer rate follow this increase). We do not count Transdniestrian part of Moldova as well as migration, which is also not included.

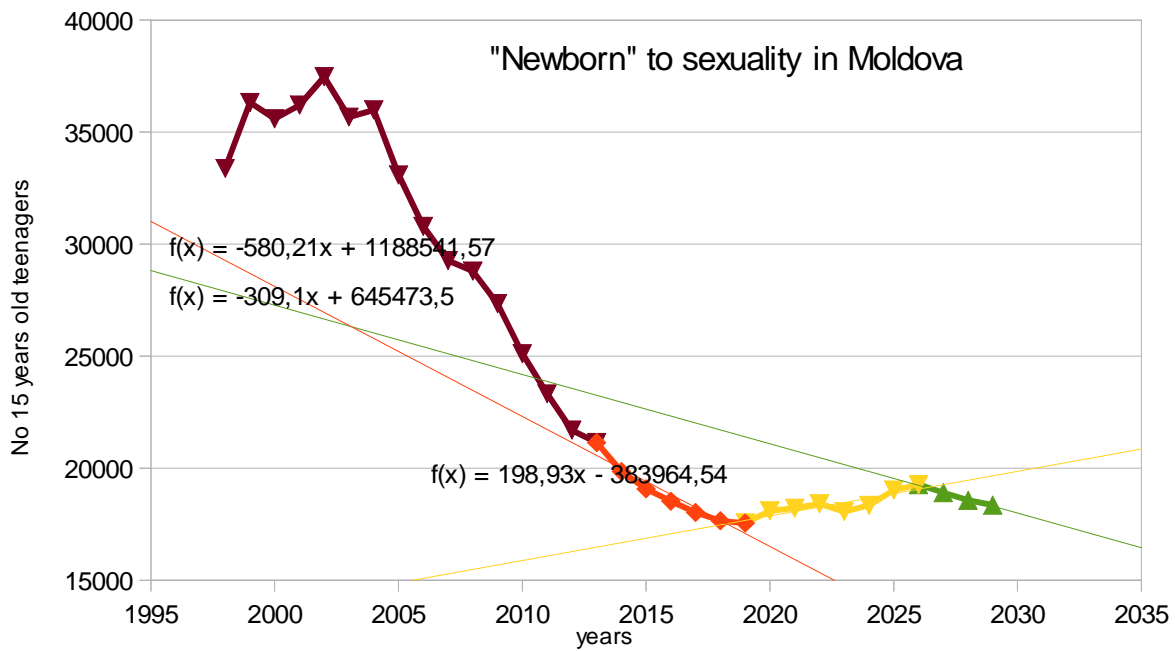

**Figure 3 Amount of 15 years old teenagers entering sexual maturity age. Historical data and projection based on population structure in 2014**

## Sexuality

First intercourse women age in Moldova (17-19) is significantly higher than in other Post-Communistic and also Western Europe countries. There are no sufficient information about partner exchange rate of adults, but some estimates as divorce patterns do not differ significantly from neighboring countries like Ukraine or Romania. We could assume, that outside of teenagers cohort, sexuality patterns will follow the same distribution as in neighboring countries. Moreover, typical for post-communistic countries liberalization of sexual life in the breakage of XX/XXI century, also took place, so we assume observing increase of sexuality. It affects mostly youngest, most susceptible to influence cohort. It start from 1991 - political transformation and beginning of social norm change in Moldova (second demographic transition or postmodernist revolution). We do not know exactly how fast such an increase is, so we will test few scenarios of it.

## Screening

Opportunistic screening was a rule in Moldova till 2011. Since 2011 Moldova introduced very novel and wide national program of common screening (women above 20 years old every 2 years). However, based on studies from 2013 around 70% adult women were never tested in their life, but more than 100 000 tests were done the same year (1/10 of targeted population). At this point screening features shows up, that young, rich, educated women from Kishinev (who are in risk group for HPV infection) screen every two years, but old, poor, less educated woman from countryside (who are in risk group for cancer development) do not do that at all.

## Vaccination

There is no public vaccination program as well, there is no money for that. There was pilot project in 2012 financed by USAIDS, but based on non-governmental studies less than 5% of targeted girls were vaccinated (amount of vaccines given to Moldova suppose to cover 75% of targeted group).

## Brief model mathematical description

Whole model description, parameters and sensitivity analysis will be further provided in Supplementary information.

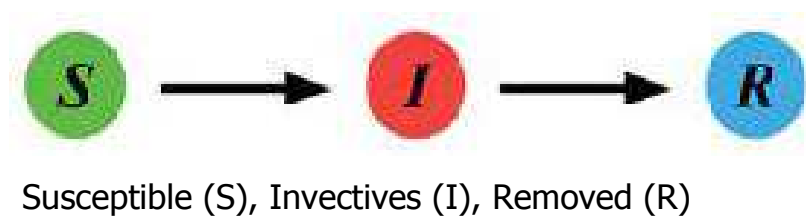

Susceptible (S), Invectives (I), Removed (R)

**Figure 4 General type of modeling infectious diseases**

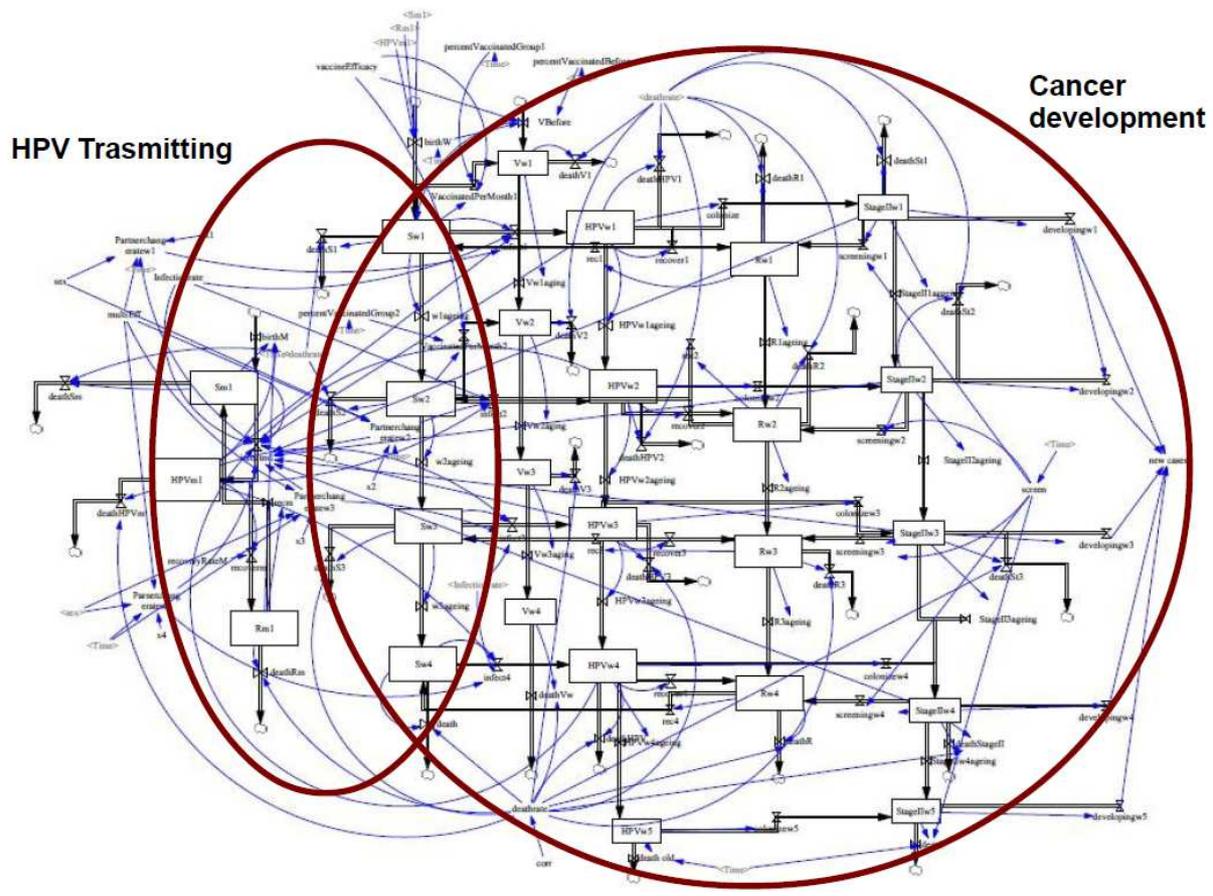

Figure 5 Overview over whole real model for imagination of complexity

**Spread of HPV  &  Cancer development  &  living Moldavian society**

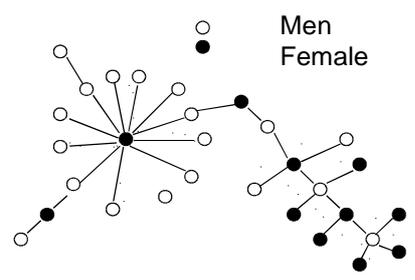

Men
Female

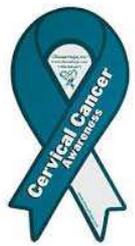

**Model**

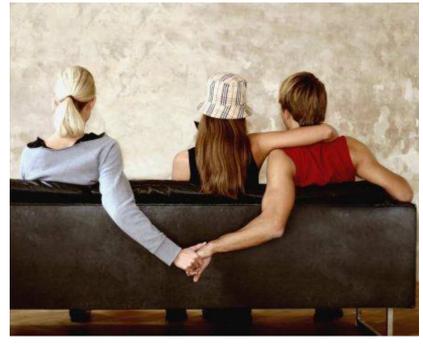

Figure 6 Model of Cervical cancer; for control: a combination of vaccination and screening; and scenarios: demographic change, increase of sexuality, screening freq

# Model preliminary results for HPV transmission and cancer development

According to provided information, model for Moldova was proposed. It's set of deterministic differential equations (implemented in Vensim). Stochasticity was introduced in sexual partner change rates. The model has aggregated the most important paths of infection for the most important pathogens. Main object is (Human Papilloma Virus, HPV) associated also with the occurrence of cervical. This is pathogen, with which has been already carefully analyzed, because its epidemiology has been widely described and modeled in recent years. We would like prepare cost/benefit analyze for different vaccination strategies, various screening programs and preventive programs (using condoms) for Moldova, based on its own demography and sexual behavior. We used data since 1998 to adjust model parameter and we project till 2031. We use mathematical and sociological concepts within complex system methodology. Mathematical modeling of infectious diseases transmitted by sexual contacts (e.g., HPV) is increasingly being used to determine the impact of possible interventions (there are dozens of such studies in literature). The most unknown parameter is the sexuality distribution and its increase. That increase was introduced as a variable. Screening efficient intervals were also implemented as a variable.

**Table 2 Model scenarios labeling**

| Levels | Sexuality | Screening frequencies |
|---|---|---|
| 0 | 10% linear increase in whole population<br>No additional sexuality increase in young cohorts (1,2) | Obtaining 10 years effective interval |
| 1 | 10% linear increase in whole population<br>Additional 10% of increase in young cohorts (1,2) | 5 years effective interval |
| 2 | 10% linear increase in whole population<br>Additional 20% of increase in young cohort (1,2) | 3 years effective interval |

Age Groups in model:
1) 15-19 (initiation of sexual live)
2) 20-24 (most active sexual group)
3) 25-34 (stabilization of sexual live)
4) 35-64 (sexual stagnation and stronger susceptibility to cancer)
5) >64 (no sexuality and cancer development)

## Incidence

The yearly incidence rates is the goal of our project. Initial condition and some model parameter were adjust to obtain empirical shape (Figure 7).

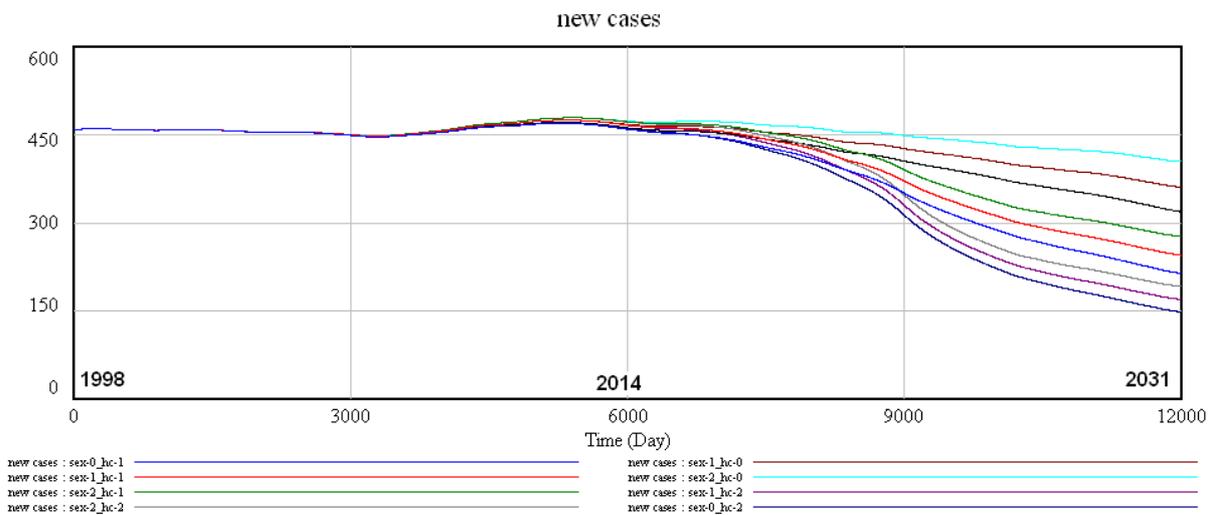

**Figure 7** Incidence of new cancer case estimated for historical data (1998-2012) and possible projections till 2031

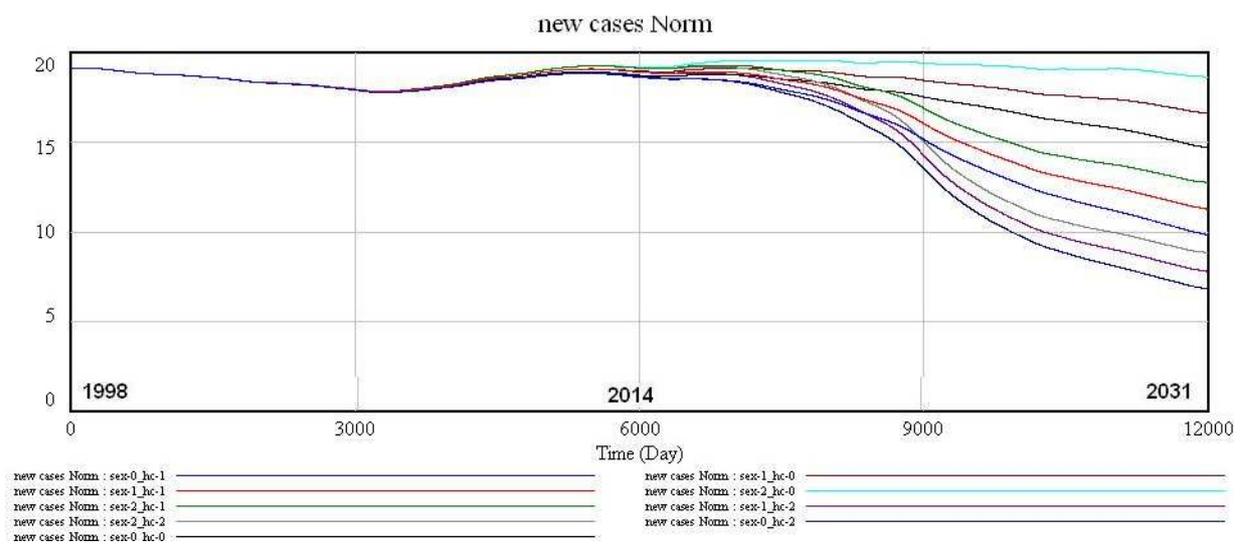

**Figure 8** Incidence rates per 100k women (age 15-65) of new cancer case estimated for historical data (1998-2012) and possible projections till 2031

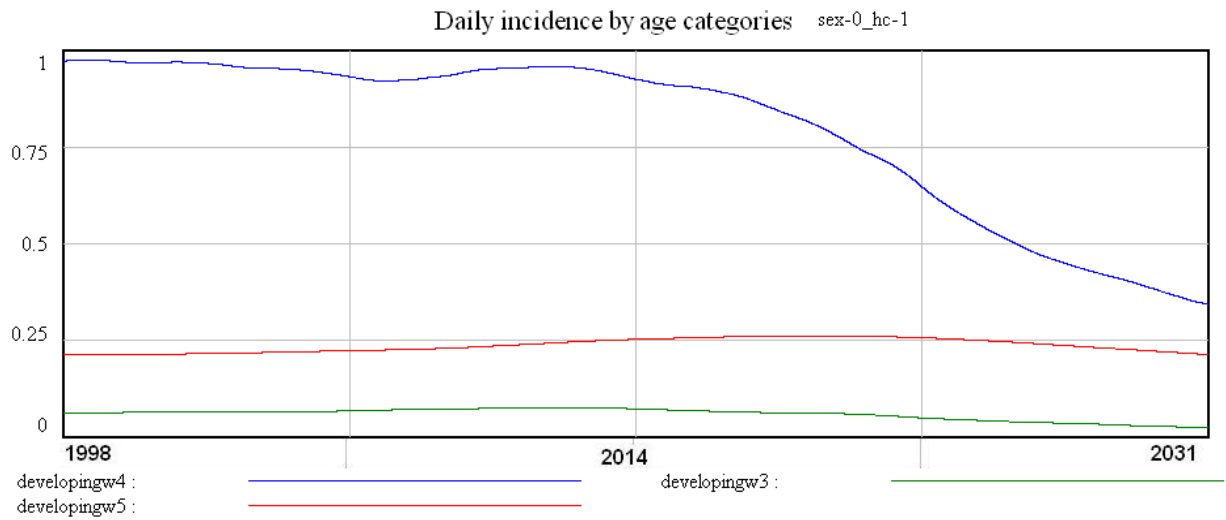

**Figure 9 Developing cancer rates for different age groups**

## Demographics

In our model Moldovan sexual active population (working age) was increasing till year 2014, and would be decreasing after. It has been also aging since around 2005.

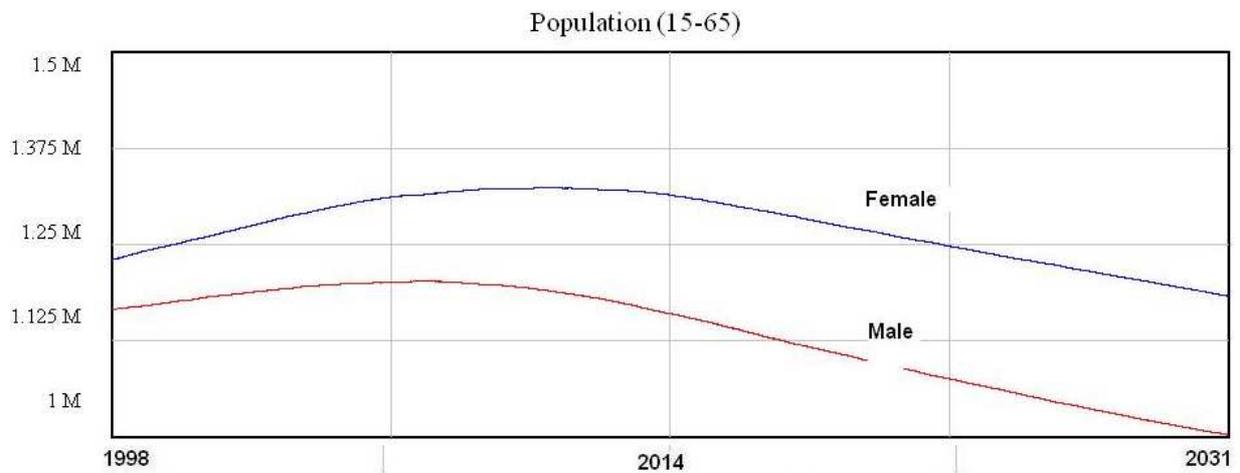

**Figure 10 Population and gender structure of Moldova (projection)**

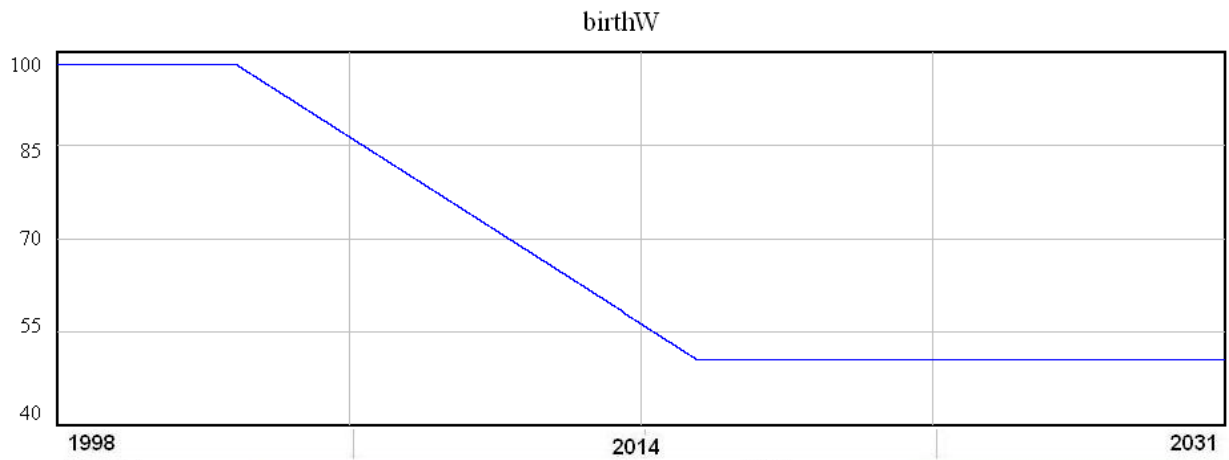

Figure 11 Entering 15 years old rates per day for girls. Compare with (with Figure 3)

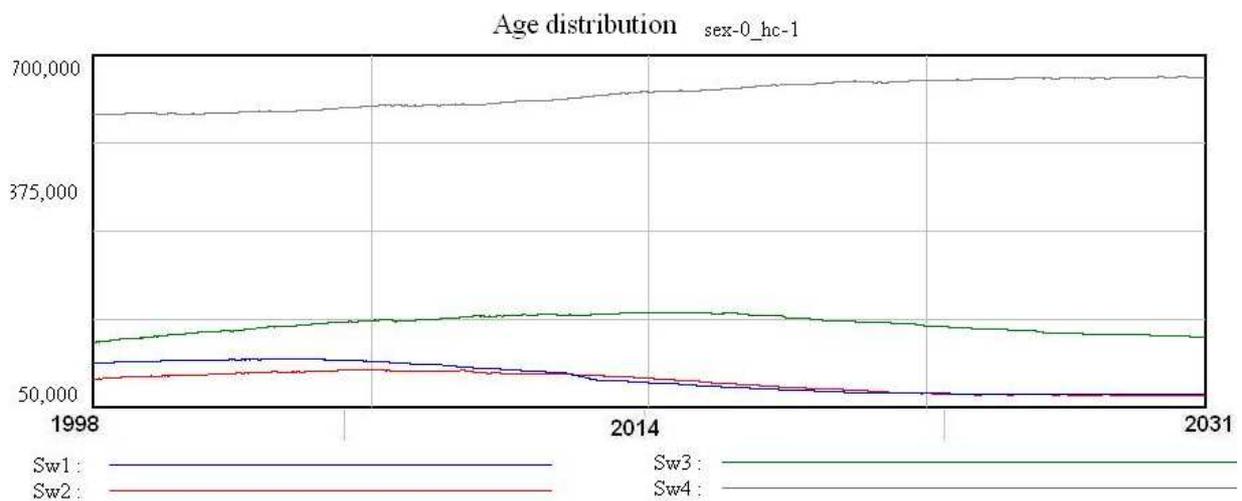

Figure 12 Change of Moldovan women age distribution with time

## Sexuality

We model sexuality starting from known form international studies distribution of sexuality. We assume intensively of partner change is on average 60% of EEA intensively (on the similar level as in Ukraine and Romania). Additionally, the sexuality of the youngest cohort is reduced next 30% (because of late first intercourse age). We try 3 levels of sexuality increase: 0-low, 1-medium, 2- high (Table 2). We assume random mixing between rural an urban populations, as well as random mixing between whole male population and different woman age-cohorts.

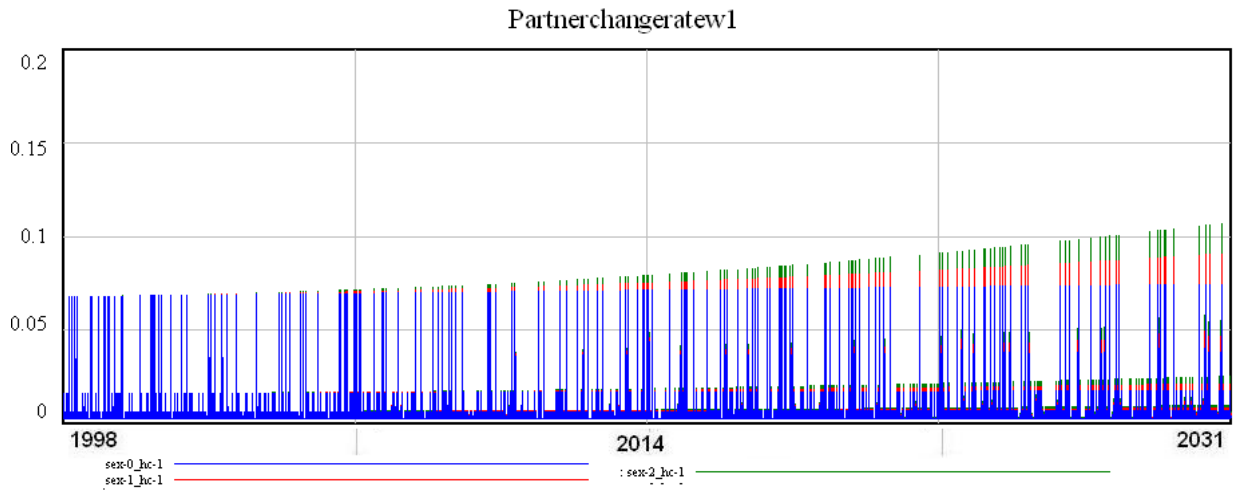

Figure 13 Sexuality increase within the youngest cohort

## Screening

Change of screening procedures from opportunistic to regular is represented in effective screening frequencies. We try 3 levels of frequency decrease: 0-low, 1-medium, 2- high. The beginning of introducing national screening program produce huge positive test results, because women, who were colonized and never tested before are captured (treatment can be applied automatically). Having experience in similar screening program in central Europe, positive test results decease from initial values (just after introducing national program) above 10% till below 5% few years later. We assume, that standard cytology test (the cheapest) will be used. It has good sensitivity and specificity for persistent colonized women and some HPV positive are also captured. Estimated overall cost of one test is 5 EUR, while cost of cancer treatment is around 1000 EUR (without social cost of it).

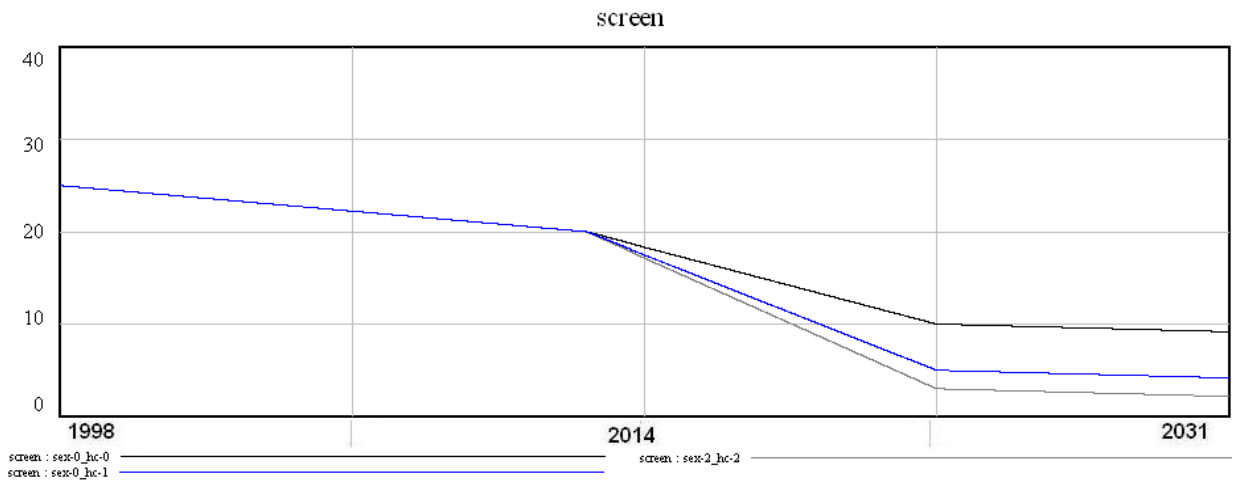

Figure 14 Screening frequencies changes

## Vaccination

We assume small opportunistic vaccination (with increase) and pilot program in 2012. We assume, that up to 5% of girls will be voluntary vaccinated.

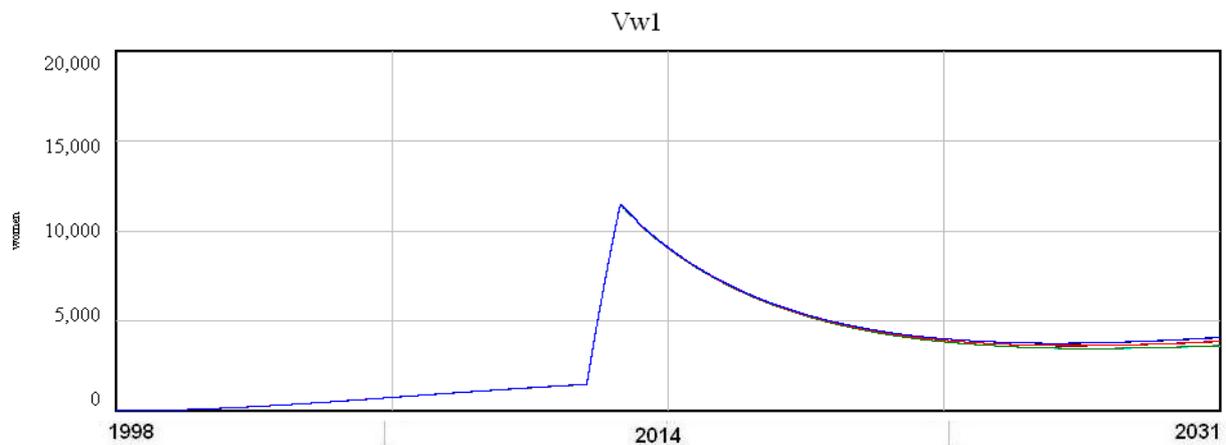

**Figure 15 Size of vaccinated cohort 15-19**

## Conclusions and Discussion with cost/benefits approach

Moldovan cervical cancer perspective seems to be much better, than in central/western Europe countries, because of relatively young society. Aging would not be a big issue (at least from HPV and Cervical Cancer perspective) in next 15 years (also comparing with central/western Europe countries). In such a preliminary setup obligatory vaccination seems to not be so crucial (for none of realistic scenarios increase of cancer cases is possible) for public health as in most countries in European Union. However, screening practice could be verified in terms of efficiency, when cost/benefit calculation would be done.

### Which screening guidelines are optimal?

Screening 20 years old girls as well as very old ladies is very costly (if all of those categories of woman test regularly it would cost ~400k EUR yearly, in current reality less than 100k EUR), but usually do not have significant effect on population (prevalence rates of cytological changes are 10 times smaller than in <25 years old cohort). Screening ladies above 65 years old is also not so efficient because the mortality caused by than other factors is much higher. They are not productive any more and aging issues would more and more problematic in Moldova in next decades.

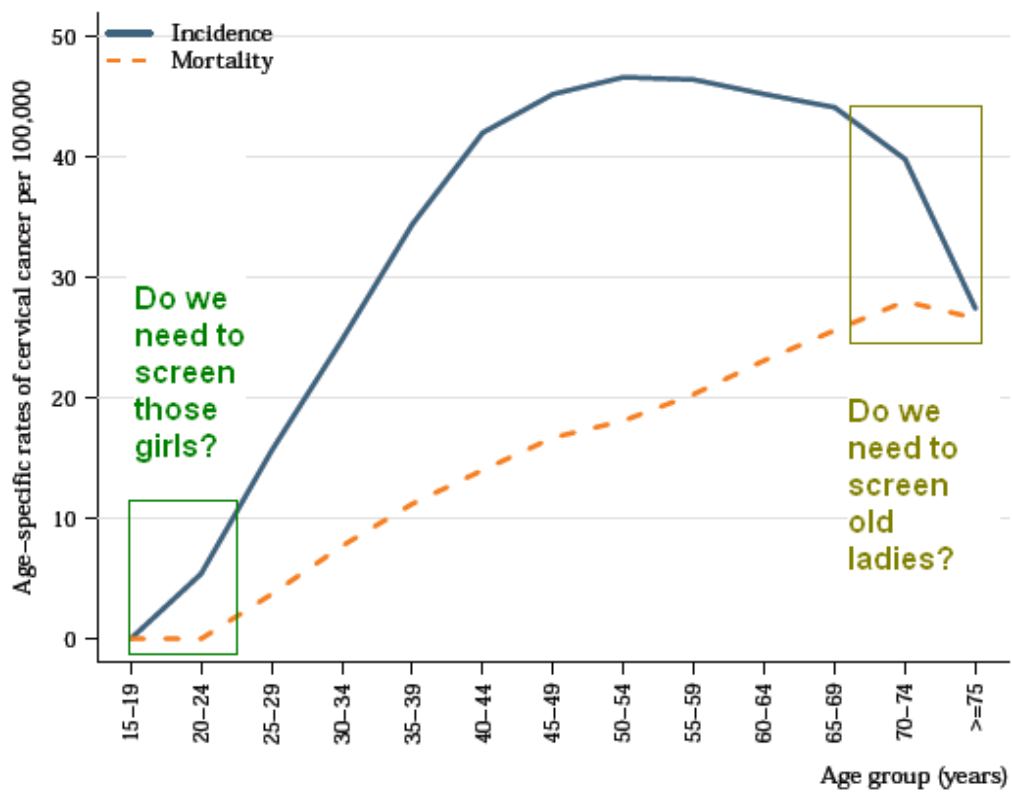

Figure 16 Mortality and incidence patterns comparison (based on https://www.hpvcentre.net)

Moreover, public heath system in Moldova has no capacity to couple with official screening program, where more than 700k test suppose to be done yearly (currently it's 100k and in many sites it seems to be already overloaded). It's more then 3.5 M EUR yearly according to this theoretical program. We are assuming 1.5k EUR standard treatment cost of cervical cancer with additional (in case of younger women) 1k EUR socio/medical costs (for woman below 65 age treatment is more expensive and state/society pays bills like sick leave, children care, lost of tax incomes). We examine overall cancer costs (Figure 17) for actual screening program +20 and (scr2) and 25-64 (scr1) where time interval frequencies are represented by health care efficiency levels hc1 (medium – 5 year efficient intervals) and level hc2 (high – 3 year efficient intervals not obtained yet even in developed countries). The saving perspective in 10-15 years would in range 100-300k EUR yearly. We also observe strange pick in cost projections (Figure 17, Figure 19) around year 2020. It's consequence of high velocity of screening interval adjusting to the national guidelines in the society. This is the point, when costs of screening exit benefits coming from reducing cancer cases. For example if we decrease effective screening intervals from every 12 years every 6 years, the cost of screening will rise twice. However, 3 years additional intervals decrease up to 3 years, makes program 4 times more expensive. This is known in economics as Laffer effect. Screening cost will overcome since 2020-25 whole others costs of cervical cancer issue for most of scenarios.

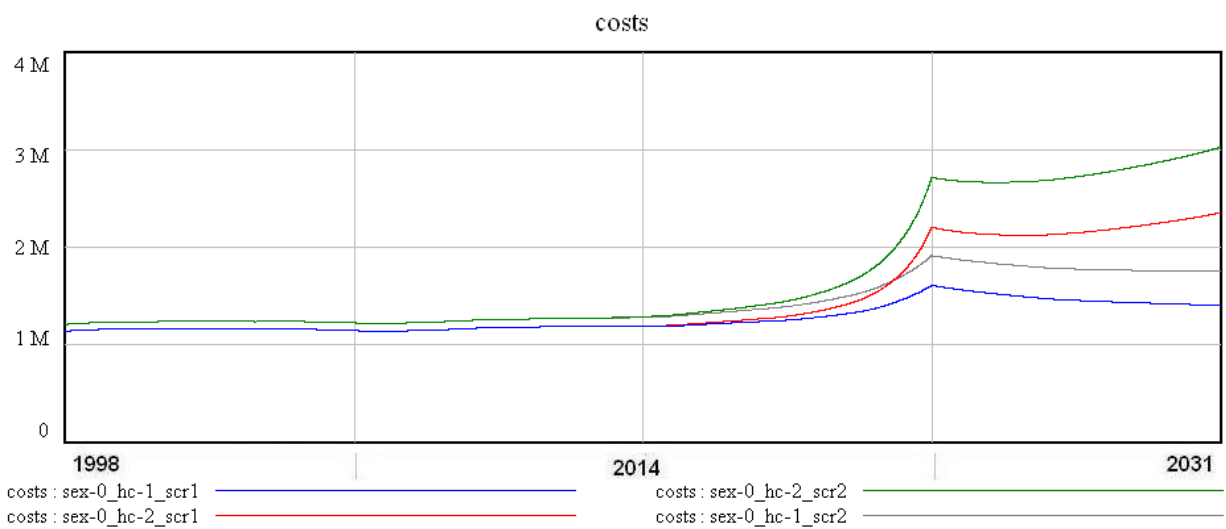

**Figure 17 Overall costs in EUR of cervical cancer (screening and treatment costs). Scr1 – optimal screening rules, Scr2-actual screening rules**

To get rid of those strange sharp picks, we could change screening function to more smooth (without sharp picks) and get more traditionally Laffel effect.

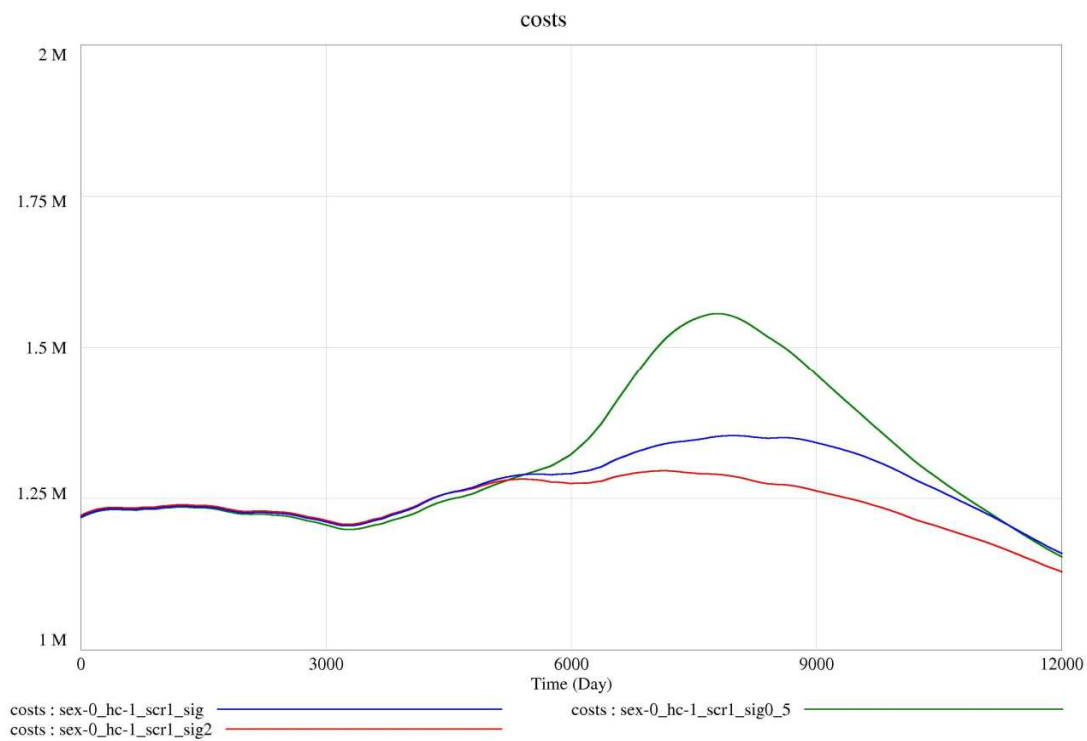

**Figure 18 Costs after smoothing procedure on screening frequency change**

# Sexual education and targeted vaccination

There is also problem with estimating sexuality risk factors. We assume, that infectivity is higher than in EEA states, because of less condom prevalence use. For viral infection like HIV, proper condom use reduce transmission probability per intercourse more than 10 times (for HPV not so many study were done, but it's shown that reduction is at least more than 2 times). Let consider intervention to teenagers age 15-19. The goal is to catch potential the most sexual active cohort (base on sociological studies according to first intercourse time, etc.) and vaccinate only 15-20% of girls with cheapest two dose vaccine (100EUR). Moreover, let introduce official sexual education (with convincing to condom use) with marginal costs (like 5EUR per each girl). In longer terms (like 30 years), that could be optimal national outcome from cancer incidence point (Figure 20) of view and the cost is very high only at the beginning of vaccination program introduction.

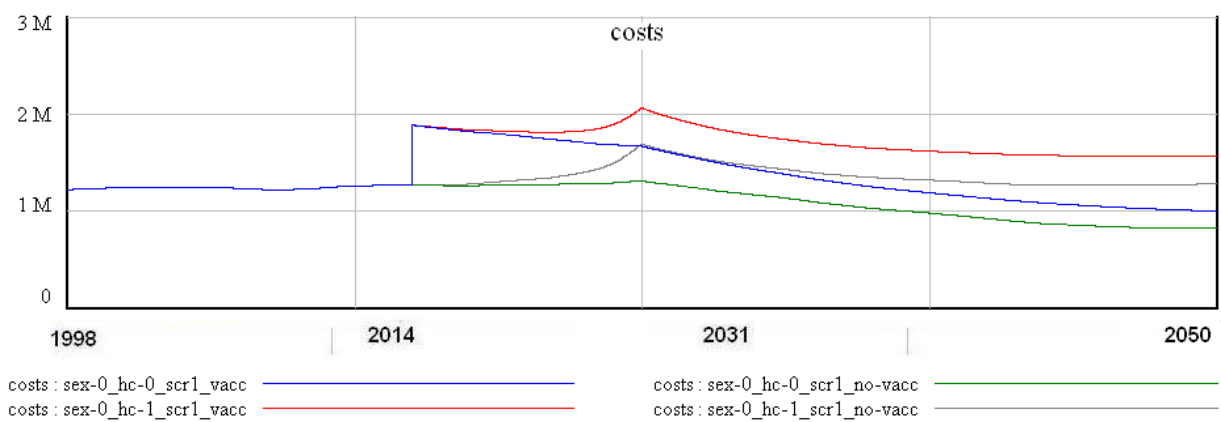

Figure 19 Cost/Benefit analysis of preferential vaccination

It seems, that targeted vaccination has similar costs in long perspective like frequent screening (compare total cost of scenarios hc1_no-vacc with hc0_vacc Figure 19) and population effect are also similar (compare number of new cases of scenarios hc1_no-vacc with hc0_vacc Figure 20). Thus, cancer treatment costs would be increasing with time, and vaccine suppose to be decreasing (unfortunately, it's opposite at this moment), so vaccination could be even more profitable.

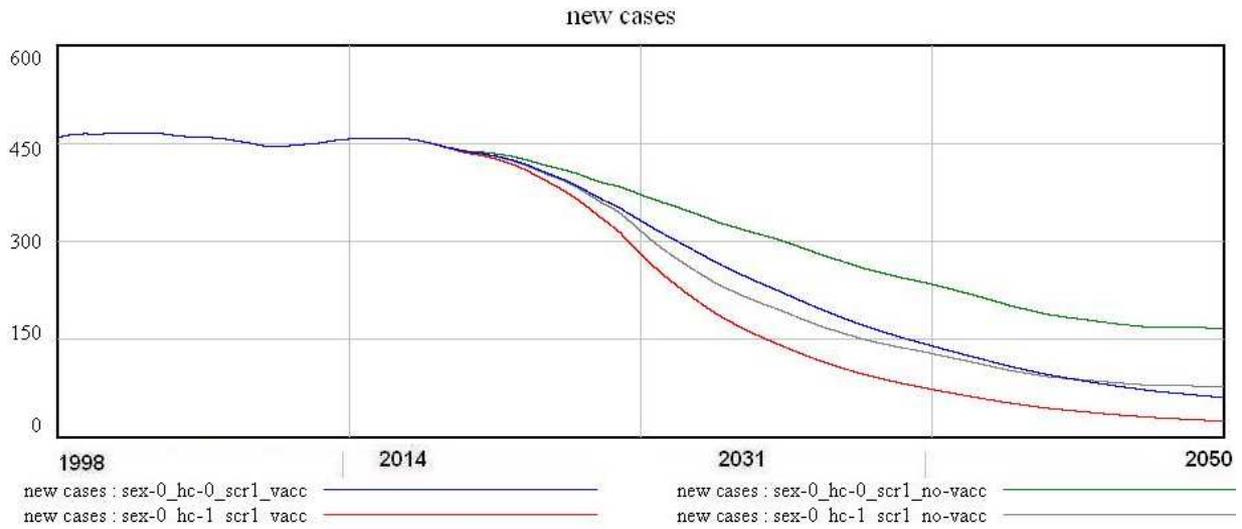

**Figure 20 Long-term new cases projections**

Despite of closed discussion about optimal HPV vaccination guidelines for developed countries (all girls below 14 years old), problem is still open in developing countries, where cost of such a action are too high (the newest generation of 3 dose vaccine could cost even 600EUR). Targeted vaccination could be consider, because costs are similar to high frequencies screening schema with the same cancer cases projection. However, some positive side effects of vaccination as reduction of pathogen circulation in society (Figure 21) will cause decrease of other pathologies related to HPV 18/16 like genital warts and other cancer (like anal). However, screening could be limited, but mustn't be cancel, because: 1) vaccine doesn't provide 100% immunity lifelong; 2) Another 10-30% of oncogenic HPV types are not stopped by vaccine.

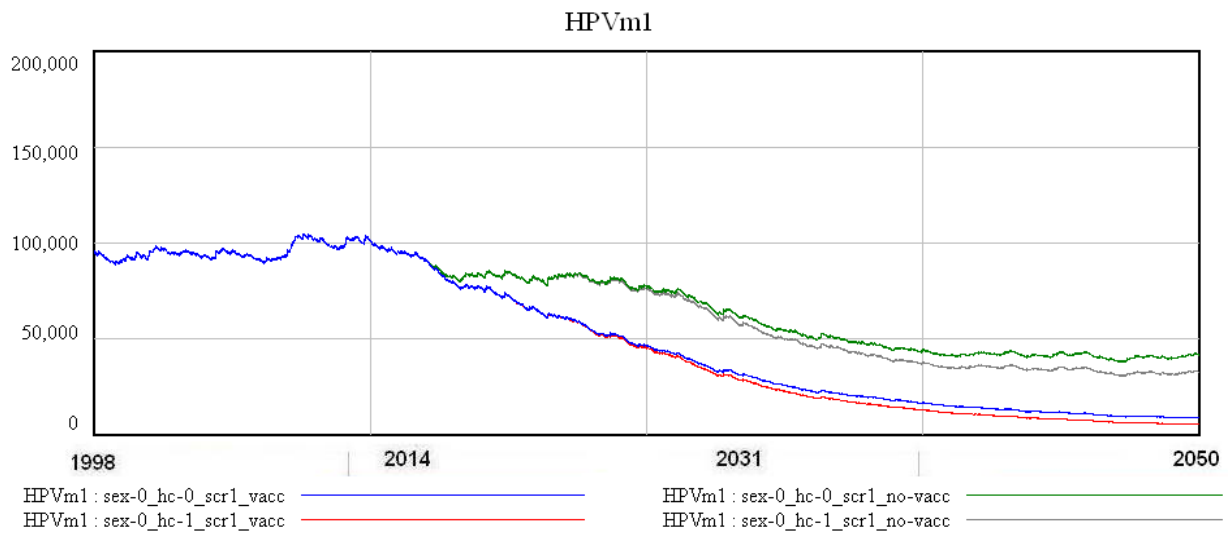

**Figure 21 Effect of vaccination to reduce prevalence of HPV in whole society (example of males)**

## Cost-effectiveness analysis

Let consider more precise changes monetary and population effects of intervention. To do so we introduce quality-adjusted life-year (QALY) - a measure of disease burden, including both the quality and the quantity of life lived. The QALY is based on the number of years of life and adjusted to health state (where 1 is a perfect health and 0 is death) that would be saved by the intervention. In previous chapters we proved, that compulsory, universal vaccination are too expensive respectively to Moldova GDP and disease will be decreasing any way - even with no interventions [Figure 7]. QALY indicates the best the benefits of intervention, because it is sensitive to current patient age (literally to current life expectancy). The intervention is cost-effective if it incremental cost per QALY yearly is below GDP per capita of given country (~2.5k EUR for Moldova) or it could be partly cost-effective if it's below 3*GDP (~7.5k EUR for Moldova).

Firstly, we compare screening scenarios. Here, we assume only two effective frequencies, medium hc1 its every 5 years and high hc2 – 3 years approaching those limits by linear splines (Figure 14). The most realistic scenario is hc1, because of it's difficult to convince all the women to screen regularly and this level of intensively of test is within Moldova's heath system capacity. The lost in cost of interventions between scr1 (our proposition) and scr2 (official realistic– yellow case in Figure 23) is around 300k EUR, and for total costs (including costs cancer curing) 250k EUR [Figure 23, Figure 22]. The gain in QALY is only 15 [Figure 23, Figure 22]. The Incremental QALY between scr1 and scr2 is 20k EUR per QALY, which is not cost-effective at all. If we compare theoretical (hc2 orange case in Figure 23) official guides with realistic official guides (hc1 yellow case in Figure 23), Incremental QALY is 1.5M/300=50k EUR per QALY. Beside hc2 is relatively good form population, it's too costly for the society.

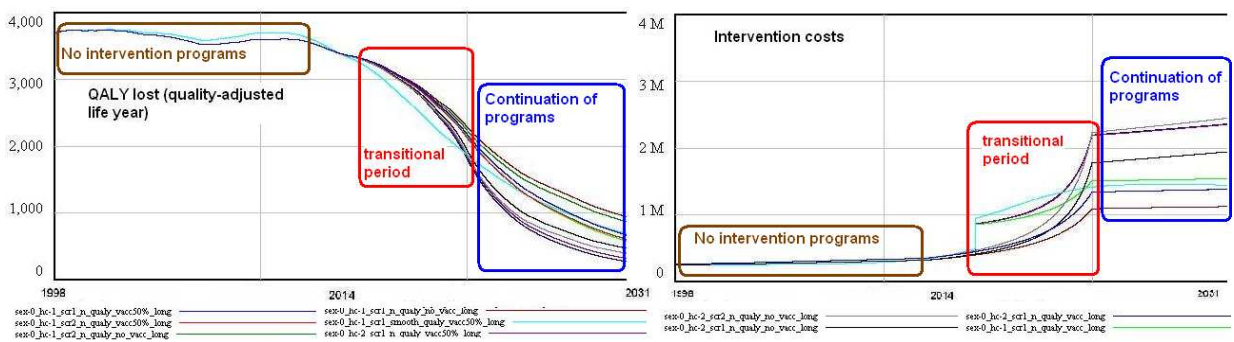

Figure 22 Past and short term analysis of main interventions of only cruel factors as QALY and Intervention cost

Another issue are targeted vaccinations. Recent studies have shown that the main safety precaution with respect to cervical cancer is going to be a combination of vaccination and screening (it's already showed that optimal is scr1). There is still open questions about vaccination, while we already proved, that all converged vaccinations are not needed and too expensive. Here we propose targeted vaccination for 20% potentially most sexually active. We choose, the cheapest 2 dose vaccine at age of 15. Within selection process, we must survey all girls of given age and we use this for sexual education (proper using condom reduces probability of infections by 50-99% dependence of source). For purpose of this research, we assume, that we will reach our targeted group in 75% (category vacc). Additionally, we also show results for 50% efficiency (category vacc50%). In vaccine efficiency analyses, we must look at longer time perspective and extend time window to 2050 for this purpose (even if we are not so sure about stability of model parameters and assumptions in such a long perspective). Vaccination (with surveys program) strategy introduce cost of 250-300k EUR yearly. However, it's in the beginning the most expensive and not the most efficient around 2020 its become cheaper than official national program (orange indicator) and ours (red indicator) from 2035 is more effective in terms of QALY [Figure 22]. By introducing this program, country saves in perspective 20-35 years 1.5M EUR according to official program with even slightly better QALY score [Figure 23]. Moreover, comparing to realistic national program (yellow indicator), with almost the same budget around 300QALY yearly will be saved.

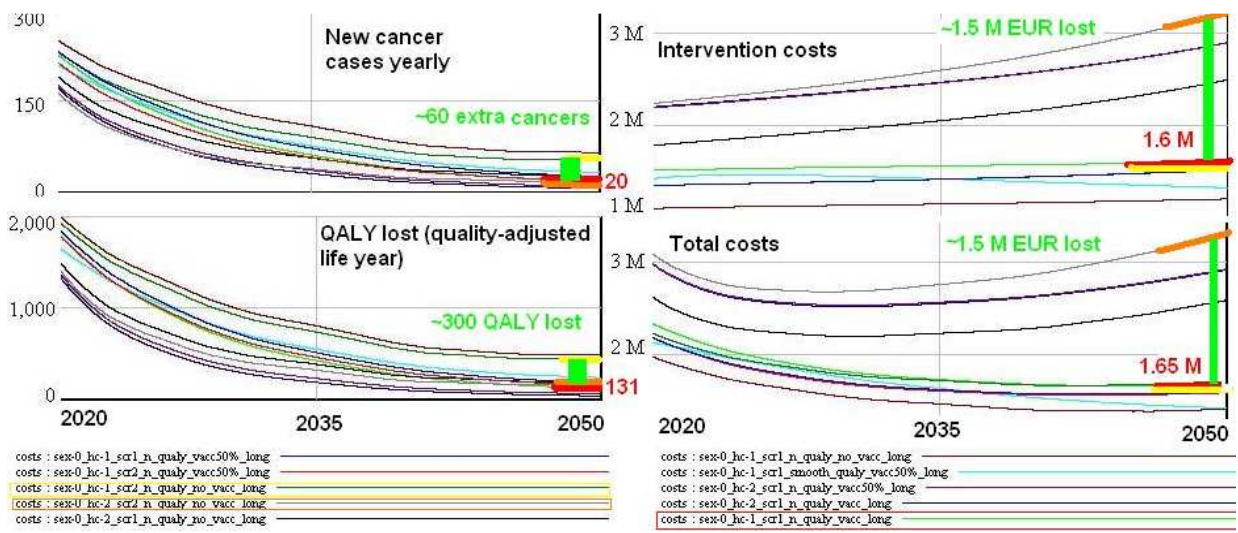

Figure 23 Long term cost-effectiveness analysis of main interventions. Variables (monetary and life) per year. Additionally, we underline 3 scenarios: our proposition (red indicator), official national program (orange indicator), realistic realization of national program (yellow indicator)

## Recommendations

Moldovan cervical cancer perspective looks much better, than in central/western Europe countries, because of relatively young society. In our setup, obligatory vaccination seems to not be so crucial (for none of realistic scenarios increase of cancer cases is possible) for public health, as in most countries in European Union.

However, screening practice could be verified in terms of efficiency and cost/benefit/effectiveness calculation must be done. Moldova has no capacity to apply official screening (all women >20 years old) program (it cost more than 3M EUR yearly according to this theoretical program) called scr2. Here more than 700k test suppose to be done yearly (currently it's 100k and in many sites it seems to be already overloaded). We propose the optimal screening guidelines (woman 25-64 old) from 2030 year with prevention cost 5-12k EUR per QALY, which could provide saving perspective in range 150k-300k EUR yearly in respect to official program.

We reopen re-open discussion about vaccination guidelines in low-income countries (as Moldova), where cost of wide action are too high. Targeted vaccination could be also consider, because costs are similar to high frequencies screening schema with the same cancer cases projection. We found that vaccination of only 20% girls chosen by surveys, will already show big difference in cancer long terms perspectives. We present optimal strategy: combination of our screening program and ours vaccination program with prevention cost 4-6k EUR per QALY from 2030. Moreover, some positive side effects of vaccination as reduction of pathogen circulation in society, will cause decrease of other pathologies related to HPV like genital warts and other cancers.

# Supplementary information

We model part of Moldovan population which is sexually active (15-64 years old). We assume a temporal naturally acquired immunity and lifelong vaccine acquired immunity, with given efficacy. Male population is divided in 3 stages: Susceptible (Sm), Infectious (HPVm), Recovered (Rm) and population is dynamic due to natural birth and death rates. For women, apart from Sw, HPVw, Rw, additional stages Long-term colonized (StageIIw), Vaccinated (Vw) and Having cancer (Cancer), are allowed and all except Cancer are divided into different age groups. We also introduced changing of society by aging, birth and death (which was set to affect older groups much more than younger ones). We assume that around the age of 65 people are not changing partners, so death rate means not natural death, but rather removing from sexually active society. We are still tracing women above 65, who were already infected (in age HPV or StageII) and can develop the cancer. We also allow dying in sexually active lifespan due to other cause, with age-dependent death rate. Birth, which is actually to be understood as turning 15 years and potentially beginning a sexual life is interpolated from register data for time up until 2014 (now) and extrapolated for near future. In our model, Moldovan society is slowly aging in waves (as it seems to be happening in reality, due to specific demographic structure). Medical properties of cancer developing are known to be age-dependent as well as sexual activity, so we decided to choose age groups with respect to data form reports. On the other hand, cancer development properties were estimated for completely different intervals. As a trade-off of having the smallest numbers of groups, but caching main difference in behavior (Age groups 1,2,3,4, +).

Transition of disease can take place due to sexual contact with given probability. With respect to Finish data set, individuals in our model have a randomly chosen number of new contacts each day, which follows the empirical distribution for given age groups from the EEA / surveys. Stochasticity was introduced not at the level of individuals, but at the level of subgroup (limitation of Vensim). One can understand that all people in a given subgroup can be very active or passive, at different iterations. We introduce change of sexuality as a linear increase of newborn's sexuality (youth at age 15 coming into the system) during 50 years of simulation sexuality level). In the older age categories sexuality increase is delayed, and agents aging (coming from younger to older subgroup) have been modeled to bring their 'sexual liberality'.

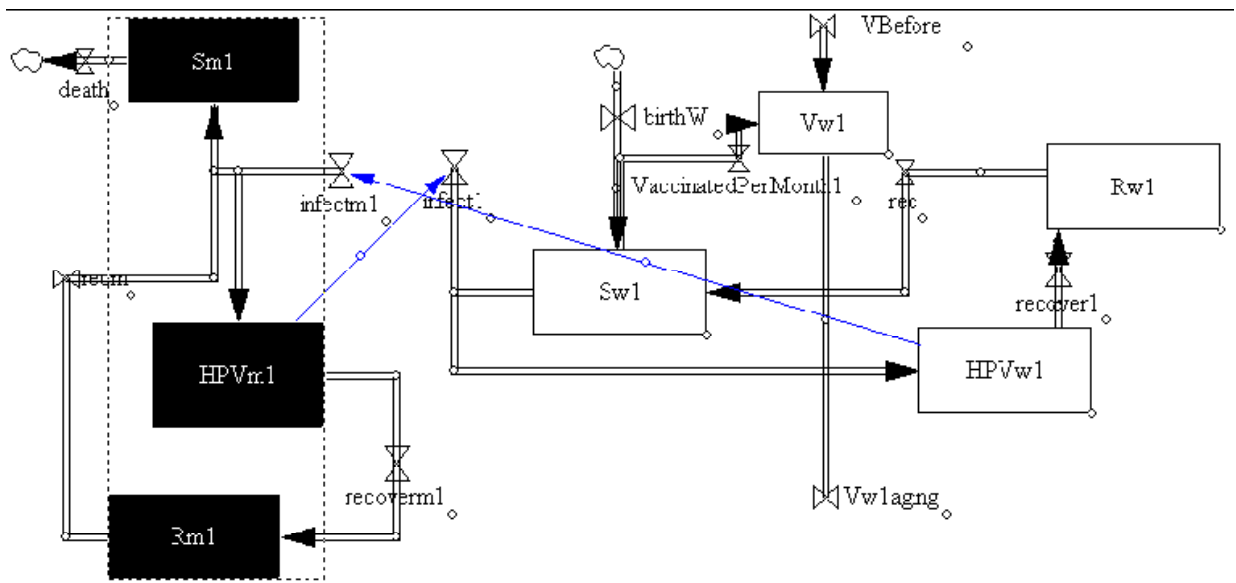

**Figure 24 Simplified spread of HPV by heterosexual contacts (m –for man, w – for women). Society is 'alive' new persons got birth, some die, some are aging**

Agents get temporal immunity due to natural recovery (and are moved from HPV or StageII to R), but after some time they lose immunity and became Susceptible (S). Moreover, recovering from HPV does not protect against new infections forever, even for the same strain, and other modelers assume immunity period of 5 or 10 years.

Once a woman is infected, she can either recover or be "permanently" colonized. From colonized state, she can be screened according to the program, or for other reasons, and find out about the infection. We assume that disease is curable in 100% if treated at this stage. Those women at the stage StageII who haven't been screened have a risk of developing cancer. Once a woman has acquired cancer, she can either survive or die, with given probabilities.

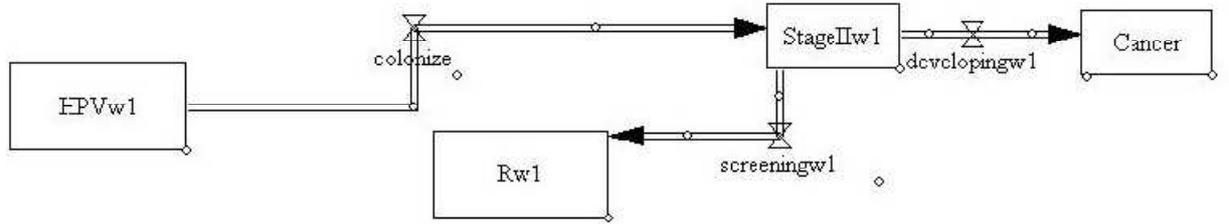

**Figure 25 Cancer developing path from HPV infection through persistent colonization (Stage II) to cancer**